\documentstyle[12pt]{article}
\newcommand{\ba}{\begin{array}}
\newcommand{\ea}{\end{array}}
\newcommand{\bd}{\begin{displaymath}}
\newcommand{\ed}{\end{displaymath}}
\newcommand{\be}{\begin{equation}}
\newcommand{\ee}{\end{equation}}
\newcommand{\bea}{\begin{eqnarray}}
\newcommand{\eea}{\end{eqnarray}}

\def\qb{\bar{q}}
\def\ub{\bar{u}}
\def\db{\bar{d}}
\def\cb{\bar{c}}
\def\sb{\bar{s}}

\def\to{\rightarrow}       % to

\def\ln{\mbox{$\ell n$}}   % logarithm

\def\Sum{\displaystyle\sum}
\def\eff{{\rm eff}}
\def\Heff{{\cal H}_{\rm eff}}

\def\me{{\bf m}}
\def\r{{\bf r}}

\def\eq{}

\def\bra{\langle}
\def\ket{\rangle}

\def\a{\alpha}
\def\b{\beta}
\def\g{\gamma}
\def\d{\delta}

\def\fifteen{{B^- \to \rho^- + \omega }}
\def\twentyseven{{B^-\to K^{*-} + \omega}}

\def\twentyeight{{B^-\to K^{*-} + \rho^0}}

\def\thirtynine{{B^-\to K^{*-} + K^{*0}}}

\def\xone{$B^-\rightarrow K^- + \eta$}
\def\xtwo{$B^+\rightarrow K^+ + \eta$}
\def\xthree{$B^-\rightarrow K^- + \eta'$}
\def\xfour{$B^+\rightarrow K^+ + \eta'$}

\def\xfive{$B^-\rightarrow K^- + \pi^0$}
\def\xsix{$B^+\rightarrow K^+ + \pi^0$}
\def\xseven{$B^-\rightarrow K^- + \eta_c$}
\def\xeight{$B^-\rightarrow D^-_S + D^0$}
\def\xeightb{$B^+\rightarrow D^+_S + \overline{D^0}$}
\def\xnine{$B^-\rightarrow \pi^- + \eta$}
\def\xten{$B^+\rightarrow \pi^+ + \eta$}
\def\xeleven{$B^-\rightarrow \pi^- + \eta'$}
\def\xtwelve{$B^+\rightarrow  \pi^- + \eta'$}

\def\xthirteen{$B^-\rightarrow \pi^- + \pi^0$}

\def\xfifteen{$B^-\rightarrow \pi^- + \eta_c$}
\def\xsixteen{$B^-\rightarrow D^- + D^0$}
\def\xsixteenb{$B^+\rightarrow D^+ + \overline{D^0}$}
\def\xseventeen{$B^-\rightarrow \pi^- +\overline{K^0}$}
\def\xseventeenb{$B^+\rightarrow \pi^+ + K^0$}
\def\xeighteen{$B^-\rightarrow   K^- + K^0$}
\def\xeighteenb{$B^+\rightarrow   K^+ + \overline{K^0}$}

\def\yone{$B^-\rightarrow K^- + \omega$}
\def\ytwo{$B^+\rightarrow K^+ + \omega$}
\def\ythree{$B^-\rightarrow K^- + \rho^0$}
\def\yfour{$B^+\rightarrow K^+ + \rho^0$}

\def\yfive{$B^-\rightarrow K^- + J/\psi$}
\def\ysix{$B^-\rightarrow D^{*-}_S + D^0$}
\def\ysixb{$B^+\rightarrow D^{*+}_S + \overline{D^0}$}

\def\yseven{$B^-\rightarrow K^{*-} + \eta $}
\def\yeight{$B^+\rightarrow K^{*+} + \eta $}

\def\ynine{$B^-\rightarrow K^{*-} + \eta'$}
\def\yten{$B^+\rightarrow K^{*+} + \eta'$}

\def\yeleven{$B^-\rightarrow K^{*-} + \pi^0$}
\def\ytwelve{$B^+\rightarrow  K^{*+} + \pi^0$}

\def\ythirteen{$B^-\rightarrow K^{*-} + \eta_c$}

\def\yfourteen{$B^-\rightarrow D_S^- + D^{*0}$}
\def\yfourteenb{$B^+\rightarrow D_S^+ + \overline{D^{*0}}$}
\def\yfifteen{$B^-\rightarrow \pi^- +\overline{K^{*0}}$}
\def\yfifteenb{$B^+\rightarrow \pi^+ + K^{*0}$}
\def\ysixteen{$B^-\rightarrow K^- + \phi$}
\def\ysixteenb{$B^+\rightarrow K^+ + \phi$}
\def\yseventeen{$B^-\rightarrow \rho^- + \overline{K^0}$}
\def\yseventeenb{$B^+\rightarrow \rho^+ + K^0$}

\def\zone{$B^-\rightarrow \rho^- + \eta$}
\def\ztwo{$B^+\rightarrow \rho^+ + \eta$}
\def\zthree{$B^-\rightarrow \rho^- + \eta'$}
\def\zfour{$B^+\rightarrow  \rho^+ + \eta'$}

\def\zfive{$B^-\rightarrow \rho^- + \pi^0$}
\def\zsix{$B^+\rightarrow  \rho^+ + \pi^0$}

\def\zseven{$B^-\rightarrow \pi^- + J/\psi $}
\def\zeight{$B^-\rightarrow  D^{*-} + D^0$}
\def\zeightb{$B^+\rightarrow D^{*+} + \overline{D^0}$}

\def\znine{$B^-\rightarrow \pi^- + \omega$}
\def\zten{$B^+\rightarrow  \pi^+ + \omega$}

\def\zeleven{$B^-\rightarrow \pi^- + \rho^0$}
\def\ztwelve{$B^+\rightarrow \pi^+ + \rho^0$}

\def\zthirteen{$B^-\rightarrow \rho^- + \eta_c$}

\def\zfourteen{$B^-\rightarrow  D^- + D^{0*}$}
\def\zfourteenb{$B^+\rightarrow D^+ + \overline{D^{0*}}$}
\def\zfifteen{$B^-\rightarrow  \pi^- + \phi$}

\def\zsixteen{$B^-\rightarrow  K^{*-} + K^0$}
\def\zsixteenb{$B^+\rightarrow K^{*+} + \overline{K^0}$}
\def\zseventeen{$B^-\rightarrow K^{-} + K^{0*}$}
\def\zseventeenb{$B^+\rightarrow K^{+} + \overline{K^{0*}}$}

\begin{document}
\vspace{-7cm}\begin{flushright} DESY 94-170\\
                                September 1994\\[1cm]
             \end{flushright}
\begin{center}
{\bf \Large
CP Violation and Strong Phases from Penguins in $\bf B^{\pm}\rightarrow PP$
and $\bf B^{\pm}\rightarrow VP$ Decays
}
\end{center}
\vspace{1cm}
\begin{center}
    G.\ Kramer$^a$,
    W.\ F.\ Palmer$^b$\footnote{Supported in part by the US
            Department of Energy under contract DOE/ER/01545-605.},
    H.\ Simma$^c$\\
      \vspace{0.3cm}
        $^a$II. Institut f\"ur Theoretische Physik\footnote{Supported by
            Bundesministerium f\"ur Forschung und Technologie,
            05\,6\,HH\,93P(5), Bonn, Germany and EEC Program
            "Human Capital and Mobility" Network "Physics at High Energy
            Colliders" CHRX-CT93-0357 (DG 12 COMA)}\\
            der Universit\"at Hamburg,\\
        D--22761 Hamburg, Germany\\
  \vspace{0.3cm}
        $^b$Department of Physics, The Ohio State University, \\
        Columbus, Ohio 43210, USA\\
  \vspace{0.3cm}
        $^c$Deutsches Elektronen Synchrotron DESY\\
        D--22603 Hamburg, Germany\\
  \end{center}
  \vspace{1cm}

\noindent {\bf Abstract}\\
\parbox[t]{\textwidth}{
We calculate direct CP-violating rate asymmetries in charged
$B\to PP$ and
$B\to VP$ decays arising from the interference of amplitudes with different
strong and CKM phases. The perturbative strong phases develop at order
$\alpha_s$ from absorptive parts of one-loop matrix elements of the
next-to-leading logarithm corrected effective Hamiltonian. CPT
constraints are maintained.  Based on this model, we find that
partial rate asymmetries between charge conjugate $B^{\pm}$ decays
can be as high as 20\% for certain channels with branching
ratios in the $10^{-6}$ range. Because the $c\bar{c}$ threshold lies so
close to the physical momentum scale, the asymmetries depend
sensitively on the model assumptions used to evaluate the imaginary
parts of the matrix elements, in particular, on the internal momentum
transfer. The charge asymmetries of partial rates would provide
unambiguous evidence for direct CP violation.
}
\newpage

\section{Introduction}

So far CP violation \cite{Jarl} has been detected only in processes
related to $K^0 - \bar K^0$ mixing \cite{CCFT} but considerable
efforts are being made to investigate it in $B$ decays. While the most
promising proposal for observing CP violation in the $B$ system involves the
mixing between neutral $B$ mesons \cite{bigi}, the particular interest
in decays of charged $B$ mesons lies in their possibilities
for establishing the detailed nature of CP violation.
Since charged $B$ mesons can not mix, a measurement of a CP violating
observable in these decays would be a clear sign for {\it direct} CP
violation, which has possibly been found in $K$ decays
where the measurements
of $\epsilon'/\epsilon$ now seem to favour a non-zero
value \cite{BP} which is consistent with expectations from the standard
model and a top quark mass around $150\, GeV$ \cite{Buras0}.

In non-leptonic charged $B$ decays two main categories of direct
CP-violating
observables can be investigated: First, rate asymmetries \cite{BSS,GH}, \be
a_{CP}=\frac{\Gamma(B^- \rightarrow f)-\Gamma(B^+ \rightarrow \bar{f})}
{\Gamma(B^- \rightarrow f) + \Gamma(B^+ \rightarrow \bar{f})}  \ ,\ee
and second, azimuthal angular correlations
\cite{Val,KP}.
The latter involve the decay of the $B$ meson into two vector particles
$B \rightarrow V_1 V_2$ with subsequent decays of $V_1$ and $V_2$
\cite{Val,KP}.
By analyzing the azimuthal dependence of the vector meson decay products
one can then isolate CP odd quantities.  The advantage of this method
is that the CP violating terms occur even when there are no strong phase
differences between the interfering weak amplitudes. On the other
hand the azimuthal correlations can also be present
when the CP-violating weak phase differences vanish. By measuring their
coefficients in charge conjugate $B^\pm$ decays one has the possibility
to disentangle the effects of strong and weak phases \cite{KP,KPS}.

The rate asymmetries occur even for final states with spinless particles
but require both weak {\it and} strong phase differences in
interfering amplitudes.  The weak phase differences arise from the
superposition of amplitudes from various penguin diagrams and the
usual W-exchange (if contributing).
It is clear that a significant
contribution of penguin diagrams, and hence of the CKM \cite{CKM} phase
differences, is an exceptional case and requires either the absence or
a strong CKM suppression of the tree contributions (as e.g. in charmless
$b \rightarrow s$ transitions).
The strong phase is generated by final state interactions. At the quark
level the strong interaction effects can be modeled perturbatively,
following Bander, Silverman and Soni \cite{BSS}, by the absorptive part
of penguin diagrams. There may be additional
final state interaction effects at the non-perturbative hadronic level
(soft final state interactions). These are very difficult to estimate,
but since the mass of the $B$ is far above the usual resonance region,
we expect these phase shifts to be small.

The rate asymmetries for exclusive two-body decays into pseudoscalars
are usually estimated using either the model of Bauer,
Stech and Wirbel \cite{BSW} (BSW) based on wave functions in the infinite
momentum frame, or the perturbative methods developed by Brodsky et al.
\cite{BLS}. The rate asymmetries $a_{CP}$ can be quite large
 (of the order $a_{CP} \sim 0.1)$
for some of the final states.  However, the corresponding branching
fractions of these decays are quite small, ranging from
$10^{-6}$ (estimates with the BSW model \cite{BSW}) to $10^{-7}$
(estimates with the Brodsky-Lepage model \cite{BLS,SW,Flei}).
The number of B-mesons required to resolve the CP asymmetry
experimentally (proportional to $(a_{CP}^2 \times BR)^{-1}$) is
therefore at least of the order of $10^{8}$.

Angular correlations and rate asymmetries for $B \to VV$ have been
investigated in some detail in our previous work
\cite{KPS}. There we considered all possible decay channels of
charged $B$ mesons into two vector particle final states which are
induced either by $b \rightarrow s$ or $b \rightarrow d$ transitions.
We found appreciable
rate asymmetries for the transitions $\twentyseven$, $\twentyeight $,
$\thirtynine $ and $\fifteen $ . For these decays the predicted rates
are necessarily small but large enough to be seen
in upcoming $B $ factories. These encouraging results have lead us
to consider the simpler decay channels of charged $B$'s into two
pseudoscalars (PP) and
into one vector plus one pseudoscalar particle (VP) within
the same framework as in our recent work for the two vector
final state \cite{KPS}. In $B \rightarrow PP$ and $B \rightarrow VP$
only one helicity amplitude contributes and thus no angular correlation
coefficients
are at our disposal for detecting direct CP violation. On the other
hand rate asymmetries might be easier to measure.

In order to systematically take into account the $O(\alpha_s)$ penguin
matrix elements, we base our treatment on the next-to-leading
logarithmic short distance corrections evaluated by Buras et al.
\cite{Buras1}.
We include also some pure penguin modes and give estimates of their
branching ratios. In this part there is overlap with other work
which we will mention later when we present our results.
Briefly, because we have a more complete treatment of the one-loop
matrix element, including all the CP constraints of \cite{GH,SEW}
and because we have included all $O(\alpha_s)$ penguins (including
so-called `hairpin' diagrams and pseudoscalar diagrams)
our results are more complete than earlier work.

The remainder of this paper is organized as follows. In Sect.~2 we
describe the effective weak Hamiltonian and the evaluation of the
hadronic matrix elements.
The final results for the branching ratios and rate
differences are discussed in Sect.~3. A table of formulae for the
various decay modes can be found in the appendix.

\section{The effective Hamiltonian}
In the next two subsections we follow closely our earlier work on
$B \rightarrow VV$ decays \cite{KPS}. To make the paper
self-contained we shall repeat some information already given in
\cite{KPS}. In subsection 2.3 we describe the evaluation of the
hadronic matrix elements which are relevant for $PP$ and $VP$ final states.

\subsection{Short distance QCD corrections}
For calculations of CP-violating observables it is most convenient to
exploit the unitarity of the CKM matrix and split the effective weak
Hamiltonian into two pieces, one proportional to
$v_u\equiv V_{ub}V_{us}^\ast$ (or $V_{ub}V_{ud}^\ast$ in the case of
$b\to d$ transitions) and the other one proportional to
$v_c\equiv V_{cb}V_{cs}^\ast$ (or $V_{cb}V_{cd}^\ast$ correspondingly),
\be
\Heff = 4 \frac{G_F}{\sqrt{2}} \left( v_u \Heff^{(u)} + v_c \Heff^{(c)} \right)
\ . \label{def_of_Hq} \ee
The two terms ($q = u,c$)
\be \Heff^{(q)} = \sum_i c_i(\mu) \cdot O_i^{(q)}\ , \ee
differ only by the quark content of the local operators,
and for our purposes it is sufficient to consider only
the following four-quark operators:
\be\ba{llllll}
O_1^{(q)} & = & \sb_\a \g^\mu L q_\b \cdot \qb_\b \g_\mu L b_\a \ , \ \ &
O_2^{(q)} & = & \sb_\a \g^\mu L q_\a \cdot \qb_\b \g_\mu L b_\b \ , \\
O_3 & = & \sb_\a \g^\mu L b_\a \cdot \Sum_{q'}\qb_\b'\g_\mu L q_\b'\ , \ \  &
O_4 & = & \sb_\a \g^\mu L b_\b \cdot \Sum_{q'}\qb_\b'\g_\mu L q_\a'\ , \\
O_5 & = & \sb_\a \g^\mu L b_\a \cdot \Sum_{q'}\qb_\b'\g_\mu R q_\b'\ , \ \  &
O_6 & = & \sb_\a \g^\mu L b_\b \cdot \Sum_{q'}\qb_\b'\g_\mu R q_\a'\ .
\label{operators}
\ea \ee
where $L$ and $R$ are the left- and right-handed projection operators.
The operators $O_3,\ldots,O_6$ arise from (QCD) penguin diagrams which
contribute at order $\a_s$ to the initial values of the coefficients at
$\mu \approx M_W$ \cite{Buras1}, or through operator mixing during the
renormalization group summation of short distance QCD corrections
\cite{QCD_SD}. The usual tree-level $W$-exchange corresponds to $O_2$
(with $c_2(M_W) = 1+O(\a_s)$). The renormalization group
evolution from $\mu \approx M_W$ to $\mu \approx m_b$ has been evaluated
in next-to-leading logarithmic (NLL) precision by Buras et al. \cite{Buras1}.
These authors also demonstrated how the $O(\a_s)$ renormalization scheme
dependence can be isolated in terms of a matrix $\r_{ij}$ by writing
\be c_j(\mu) = \sum_i \bar{c}_i(\mu)
 \left[ \d_{ij} - \frac{\a_s(\mu)}{4\pi} \r_{ij}\right] \ , \ee
where the coefficients $\cb_j$ are scheme independent at this order.
The numerical values for $\Lambda^{(4)}_{\overline{MS}} = 350$ MeV
(which is in accord with $\Lambda^{(5)}_{\overline{MS}}
= 240\pm 90 MeV$ from the compilation of G. Altarelli \cite{Alt}),
$m_t = 150$ GeV and $\mu = m_b = 4.8$ GeV are \cite{Buras1}
\be\ba{llllll}
\cb_1 & = & - 0.324\ , \ \ & \cb_2 & = & 1.151\ , \\
\cb_3 & = & 0.017\ , \ \ & \cb_4 & = & -0.038\ , \\
\cb_5 & = & 0.011\ , \ \ & \cb_6 & = & -0.047\ .
\ea\ee

Contributions from the color magnetic moment operator
will always be neglected in the following, because already its tree level
matrix elements are suppressed by a factor $\alpha_s/4\pi$ and it cannot
provide interesting absorptive parts in the decays considered here.

\subsection{Quark-level matrix elements}
Working consistently at NLL precision, the matrix elements of $\Heff$ are
to be treated at the one-loop level in order to cancel the scheme
dependence from the renormalization group evolution. The one-loop matrix
elements can be rewritten in terms of the tree-level matrix elements of
the effective operators:
\be \bra sq'\qb'\vert \Heff^{(q)}\vert b\ket =
\sum_{i,j} c_i(\mu)
\left[ \d_{ij} + \frac{\a_s(\mu)}{4\pi} \me_{ij}(\mu,\ldots)\right]
\bra sq'\qb'\vert O_j^{(q)}\vert b\ket^{\rm tree} \ . \label{me1}\ee
The functions $\me_{ij}$ are determined by the corresponding renormalized
one-loop diagrams and depend in general on the scale $\mu$, on the quark
masses and momenta, {\it and} on the renormalization scheme. The various
one-loop diagrams can be grouped into two classes: {\it vertex-corrections},
where a gluon connects two of the outgoing quark lines (fig.~1a),
and {\it penguin} diagrams, where a quark-antiquark line closes a loop
and emits a gluon, which itself decays finally into a quark-antiquark pair
(fig.~1b).

When expressing the rhs of \eq(\ref{me1}) in terms of the renormalization
scheme independent coefficients $\cb_i$, the effective coefficients multiplying
the matrix elements $\bra sq'\qb'\vert O_j^{(q)}\vert b\ket^{\rm tree}$
become
\be c_j^\eff \equiv \cb_j + \frac{\a_s}{4\pi} \sum_i \cb_i \cdot
\left( \me_{ij}-\r_{ij} \right) \ . \ee
The renormalization scheme dependence, which is present in $\me_{ij}$
and $\r_{ij}$, explicitly cancels in the combination $\me_{ij}-\r_{ij}$
\cite{Buras1}.

The effective coefficients $c_{1,2}^\eff$ receive contributions only from
{\it vertex-correction} diagrams, which will not be included in the following
(see discussion at the end of Section~2.3).
For a general $SU(N)$ color group the remaining effective coefficients can
be brought into the following form
\bea
c_3^\eff & = & \cb_3 - \frac{1}{2N} \frac{\a_s}{4\pi} (c_t + c_p)
     + \cdots \nonumber\\
c_4^\eff & = & \cb_4 + \frac{1}{2} \frac{\a_s}{4\pi} (c_t + c_p)
     + \cdots \nonumber\\
c_5^\eff & = & \cb_5 - \frac{1}{2N} \frac{\a_s}{4\pi} (c_t + c_p)
     + \cdots \nonumber\\
c_6^\eff & = & \cb_6 + \frac{1}{2} \frac{\a_s}{4\pi} (c_t + c_p)
    + \cdots \ , \label{c3456} \eea
where we have separated the contributions $c_t$ and $c_p$ from the ``tree''
operators $O_{1,2}$ and from the penguin operators $O_{3\cdots6}$,
respectively. The ellipses denote further contributions from
{\it vertex-correction} diagrams.

In addition to the contributions from penguin diagrams with insertions of
the tree operators $O^{(q)}_{1,2}$
\be
c_t = \cb_2 \cdot \left[\frac{10}{9}+\frac{2}{3} \ln \frac{m_q^2}{\mu^2}
- \Delta F_1\Bigl(\frac{k^2}{m_q^2}\Bigr) \right]\ , \label{ct} \ee
where $\Delta F_1$ is defined in \cite{KPS}, we have evaluated
the penguin diagrams for the matrix elements of the penguin operators:
\bea
c_p & = & \cb_3 \cdot \left[
       \frac{280}{9}
       + \frac{2}{3} \ln \frac{m_s^2}{\mu^2}
       + \frac{2}{3} \ln \frac{m_b^2}{\mu^2}
       - \Delta F_1\Bigl(\frac{k^2}{m_s^2}\Bigr)
       - \Delta F_1\Bigl(\frac{k^2}{m_b^2}\Bigr) \right] \nonumber \\
& + & (\cb_4+\cb_6)\cdot \sum_{j=u,d,s,\ldots} \left[
       \frac{10}{9}+\frac{2}{3} \ln \frac{m_j^2}{\mu^2}
      - \Delta F_1\Bigl(\frac{k^2}{m_j^2}\Bigr) \right] \ , \label{cp}
\eea
Note that the coefficients $c_i^\eff$ depend on $k^2$ and, as we shall
see later, on the $q\bar{q}$ states that are included in the sum over
the intermediate states.

\subsection{Hadronic matrix elements in the BSW model}
To take into account long distance QCD effects which build up the
hadronic final states, we follow Bauer, Stech and Wirbel \cite{BSW}:
With the help of the factorization hypothesis \cite{Fak} the three-hadron
matrix elements are split into vacuum-meson and meson-meson matrix elements
of the quark currents entering in $O_1,\ldots,O_6$. In addition, OZI
suppressed form factors and annihilation terms are neglected.
In the BSW model, the meson-meson matrix elements of the currents are
evaluated by overlap integrals of the corresponding wave functions and
the dependence on the momentum transfer (which is equal to the mass of
the factorized meson) is modeled by a single-pole ansatz.
As a first approximation, this  calculational scheme
provides a reasonable method for estimating the relative size and
phase of the tree and penguin terms that give rise to the CP-violating
signals. It is known that the BSW matrix elements do not describe fully
the existing experimental information on decays like $B \rightarrow
J/\psi + K, J/\psi + K^*$ \cite{CLEO}. But this should not matter too much
for the estimates of CP effects presented here.

Compared to the transitions $B \rightarrow VV$ treated in our earlier
work \cite{KPS} the amplitudes for $B \rightarrow PP$ and
$B \rightarrow PV$ have a simpler structure because only one helicity
state contributes.
On the other hand there are additional contributions from the $(V+A)$
penguin operator $O_6$: After Fierz reordering and factorization
it contributes in terms which involve a matrix element of the
quark-density operators between a pseudoscalar meson and the
vacuum
\be
\bra P_1 M_2\vert O_6 \vert B \ket = - 2 \Sum_q \Bigl(
\bra P_1 \vert \bar q b_L \vert 0\ket \bra M_2 \vert\sb q_R \vert B \ket +
\bra M_2 \vert \bar q b_L \vert 0\ket \bra P_1 \vert\sb q_R \vert B \ket
\Bigr) \ . \ee

Using the Dirac equation, the matrix elements entering here can be
rewritten in terms of those involving usual $(V\!-\!A)$ currents,
\be
   \bra P_1 M_2 \vert O_6 \vert B \ket =
   R[P_1,M_2] \bra P_1 M_2 \vert O_4 \vert B \ket \ ,
\label{O6_vs_O4} \ee
with
\be
R[P_1,M_2] \equiv
\frac{\pm 2M^2_{P_1}}{(m_{q1}+m_{\bar q1})(m_b \mp m_{q2})} \ .
\label{def_of_R} \ee
Here, $m_{q1}$ ($m_{\bar q1}$) and $m_{q2}$ are the current masses of the
(anti-)quark in the mesons $P_1$ and $M_2$, respectively, and the
the upper (lower) sign is for the $PP$ ($PV$) final state.
We use the quark masses $m_u=m_d= 10$~MeV, $m_s=200$~MeV, $m_c = 1.5$~GeV
and $m_b = 4.8$~GeV.

Finally, one arrives at the form
\bea \bra P_1\,M_2\vert \Heff^{(q)}\vert B\ket
& = & Z^{(q)}_1 \bra P_1 \vert j^{\mu}\vert 0 \ket
          \bra M_2 \vert j_{\mu} \vert B\ket \nonumber \\
& + & Z^{(q)}_2 \bra M_2 \vert j'^{\mu}\vert 0 \ket
          \bra P_1 \vert j'_{\mu} \vert B\ket \ ,
\label{def_of_Z12} \eea
where $j_\mu$ and $j'_\mu$ are the corresponding (neutral or charged)
$V\!-\!A$ currents. The factorization coefficients $Z^{(q)}_1$ and
$Z^{(q)}_2$ are listed in the appendix.
In terms of the form factors $F_{0,1}$ and $A_0$ for the current matrix
elements defined by BSW \cite{BSW}, this yields
\bea \bra P_1 P_2 \vert \Heff^{(q)}\vert B\ket  & = &
        Z^{(q)}_1 (M_B^2-M_2^2) f_{P_1} F_0(M_1^2)
      + Z^{(q)}_2 (M_B^2-M_1^2) f_{P_2} F_0(M_2^2) \ ,\\
    \bra P_1 V_2 \vert \Heff^{(q)}\vert B\ket  & = &
    \lambda^\frac{1}{2}(M_B^2,M_P^2,M_V^2) \left(Z^{(q)}_1 f_{P}
    A_0(M_P^2) + Z^{(q)}_2 f_{V} F_1(M_V^2)\right) \ .
\eea
$M_B$ is the mass of the decaying $B$ meson and $f_P$ ($f_V$) are the
decay constants of the pseudoscalar (vector) mesons in the final state.

Concerning how $1/N$ terms are treated in the coefficients (see
\eq(\ref{c3456}) and \eq(\ref{def_of_ai})), it is well known \cite{Stone}
that this model has problems accounting for the decays with branching
ratios which are proportional to the combination
$ \bar{c}_1 + \bar{c}_2/N$. This is due to the rather small absolute
value of this particular combination when using the short-distance QCD
corrected coefficients\footnote{An interesting analysis
  of the value of $c_1+c_2/N$ has recently been presented in
  ref.~\cite{Buras_a1} pointing out the renormalization dependence of
  the QCD short distance corrections}.
An analogous
effect is also known in nonleptonic D decays \cite{BSW}, and several
authors advocated a modified procedure to evaluate the factorized
amplitudes \cite{BSW,Buras2}: There, only terms which are dominant in
the $1/N$ expansion are taken into account.
Recently there has been much discussion in the literature concerning
these issues.  Some authors have argued that QCD sum rules validate
this procedure \cite{sum}.
We also choose this leading $1/N$ approximation in evaluating the matrix
elements of the weak Hamiltonian and we use the QCD corrected
coefficient functions $\cb_i$ given above.

The strong phase shifts are generated in our model only by the
absorptive parts (hard final state interactions) of the quark-level
matrix elements of the effective Hamiltonian. Of course, when factorizing
the hadronic matrix elements, all information on the crucial value of the
momentum transfer $k^2$ of the gluon in the penguin diagram (fig.~1b) is
lost. While there has been an attempt \cite{SW}
to model a more realistic momentum distribution by taking into account
the exchange of a hard gluon, we will use here for simplicity only a
fixed value of $k^2$. From simple two body kinematics \cite{Deshpande}
or from the investigations in ref.~\cite{SW} one expects $k^2$ to be
typically in the range
\be
\frac{m_b^2}{4}\stackrel{<}{\sim} k^2 \stackrel{<}{\sim}\frac{m_b^2}{2}\ .
\label{kk_range} \ee

The results we shall present are sensitive to $k^2$ in this range
because the $c\bar{c}$ threshold lies between these limits. Arguments
have been made that the lower limit is a more appropriate choice
\cite{SEW}. In this work we follow \cite{KPS} and choose the upper
limit for our numerical presentation in the tables. However, we have
studied the
asymmetries as a function of $k^2$ and will show the results later
in Fig. 2-4.

While the next-to-leading logarithmic precision of the effective
Hamiltonian allows one to consistently calculate all amplitudes at order
$\alpha_s$ and to include all one-loop matrix elements, some care is
necessary when evaluating CP-violating asymmetries of the decay rates.
In particular, one should make sure that the rate asymmetries for
sufficiently inclusive channels remain consistent with CPT constraints
in certain mass limits \cite{GH}. This issue has been discussed in some
detail in \cite{KPS} and we shall follow a similar procedure of
neglecting absorptive parts from flavor diagonal rescattering.

We refined here the approximation of ref.~\cite{KPS} by dropping only
those fractions of the imaginary parts of
$\Delta F_1(k^2/m_q^2)$ in (\ref{ct}) and (\ref{cp}) which correspond
to the projection of the intermediate quark state containing a
$q\bar{q}$-pair onto the final state.
%%% Of course, we account thereby
%%% only for flavour and color factors, but not for hadronization
%%% effects from wave function overlaps and long distance physics.
Moreover, no absorptive parts of vertex correction diagrams are included
in our calculations, because they are always flavour diagonal.
Of course, this approach may overestimate the CPT cancellation
of the absorptive phases.
We have explicitly checked for all decay channels that
taking into account the full imaginary parts of the penguin-like matrix
elements (\ref{ct}) and (\ref{cp}) --- including flavour-diagonal
contributions --- does not significantly change our results. For the
most sensitive cases
$B\to \pi^- \eta'$ and $B\to \rho^- \eta'$ the inclusion
of all intermediate flavours would shift the asymmetries by at most
$-10$\%, which is in these cases still less than the uncertainties
from the value of $k^2$.

As a result of our procedure, the factorization coefficients
$Z^{(q)}_{1,2}$ depend not only on whether the basic tree level operator
is of the $u\bar{u}$ or $c\bar{c}$ type, and on the effective
value of $k^2$, but also on the quark content ($u\bar{u},d\bar{d},
s\bar{s},c\bar{c}$) contributing in the one-loop penguin matrix elements.
As in \cite{KPS} we do not explicitly drop higher order terms
which arise, for instance, through interferences among (real and imaginary
parts of) the order $\a_s$ matrix elements.
However, such terms can not introduce the above mentioned inconsistencies
with CPT because the flavour-diagonal absorptive parts are discarded.

\section{Results and Discussion}

For a numerical analysis of the decay parameters and their CP-violating
effects within our model, we need to specify the CKM  matrix elements and
the current form factors. It is well known
\cite{schu} that fits for the parameters \footnote{
These coincide with the parameters $\rho$ and $\eta$ of the Wolfenstein
representation for small angles \cite{wolf1}. }
\bea
\rho & = & cos \delta_{13}\; s_{13}/(s_{12}s_{23)}\nonumber \\
\eta & = & sin \delta_{13}\; s_{13}/(s_{12}s_{23)}
\eea
of the CKM matrix depend critically on the value of the $B$-meson decay
constant $f_B$. The solution for lower $f_B$ values leads to a negative
$\rho$ while higher $f_B$ values render $\rho$ positive.

We have calculated our results for the two solutions, with the
values
\bea
\rho = 0.32\ ,\ & \ \ \eta = 0.31,\ \ \ & (f_B=250\,MeV,~
m_t = 135\pm 27\,GeV)   \nonumber \\
\rho =-0.41\ ,\ & \ \ \eta = 0.18,\ \ \ & (f_B=125\,MeV,~
m_t = 172\pm 15\,GeV)   \nonumber
\eea
from the analysis by Schmidtler and Schubert\cite{schu}.
A more recent analysis by Ali and London \cite{Ali} based on the latest
information on $V_{ub}$ yields similar results.

The results for the modes which get contributions from tree and penguin
diagrams are collected in Tab.~1, 2, 3 and 4. They are obtained with
$k^2=m_b^2/2$ as in our earlier work \cite{KPS} and for the two CKM
parametrizations which we call the $\rho > 0$ and $\rho < 0$ solution,
in the following discussion. The results
for $B \rightarrow PP$ are given in Tab.~1 ($b \rightarrow s$
transitions) and Tab.~2 ($b \rightarrow d$ transitions).
$B \rightarrow PV$ results are given in Tab.~3 ($b \rightarrow s$
transitions) and Tab.~4 ($b \rightarrow d$ transitions).
The pattern of branching ratios of the
various channels is rather similar for $B \rightarrow PP,\,
B \rightarrow PV$ and $B \rightarrow VV$ \cite{KPS}.
Compared to $B \rightarrow VV$ decays considered in \cite{KPS} the
branching ratios are similar but the asymmetries are somewhat
smaller. Nevertheless there are some interesting cases with large
enough branching ratios and asymmetries to be measurable in
currently planned machines.
Some differences between the pattern for $B \rightarrow PP$ and $VP$
and that for $B \rightarrow VV$ are due to
the contribution of the operator $O_6$ (see \eq(\ref{O6_vs_O4})).

If we look at specific decay channels the most interesting cases in
Tab.~1 are the decays \xthree ~ and \xfive. The branching ratios are of
the order of $10^{-5}$ and the asymmetries in the several percent
range. For $\rho > 0$ they are larger than for $\rho < 0$ whereas the
branching ratios show the opposite pattern. The channel \xseventeen ~
is a pure penguin transition with a moderately large branching ratio
but a small asymmetry. In Tab.~2 \,\xthirteen ~ has no asymmetry since
the penguins cancel due to isospin symmetry. Interesting cases are
\xnine ~ and \xeleven ~ with branching ratios of the order of $10^{-5}$
and asymmetries of several percent. It is interesting to note that
\xsixteen ~ has a substantOBial rate as well as an asymmetry of the order
of $2 \%$, independent of whether $\rho$ is positive or negative.
The last channel,~\xeighteen ~ is again a pure penguin transition with
large asymmetry for postive $\rho$ but with a small branching ratio.

In Tab.~3 we give the $B \rightarrow PV$ results ($b \rightarrow s$
transitions).
The cases with the biggest asymmetries have very small branching
ratios ($\approx 10^{-7}$). Interesting decays are \ynine\
%%% ($\rho >0$),
\yseven\
%%% ($\rho < 0$)
and \yeleven\ . %%% for both $\rho $ solutions.
The three last pairs of decays at the bottom of Tab.~3 are again pure
penguin modes which have reasonable branching ratios but small
asymmetries.

Tab.~4 contains the results for $B \rightarrow PV$ ($b \rightarrow d$
transitions). The last three pairs
of decays in this table are again pure penguin modes. The transition
\zfifteen\ has zero asymmetry since no (absorptive) penguin-like matrix
elements are present in the contributing factorization due to its color
structure.
The other pure penguin modes show appreciable asymmetries
for the positive $\rho $ solution. Their branching ratios are small,
however. Concerning the other decays, the channel \zthree ~ stands out
with a branching ratio of $10^{-5}$ and an asymmetry of $14 \%$
for positive $\rho$ and a branching ratio of $6\times 10^{-5}$ with an
asymmetry of $1.2\%$ for the negative $\rho$ solution.
Other interesting cases are \zone ~, \zfive ~, \znine ~ and \zeleven ~
with significant asymmetries. For $\rho < 0$ the asymmetries are usually
smaller than for $\rho > 0$ but then the branching ratios are larger.

Of course, these results can not be considered definitive since we had
to make a series of model assumptions namely factorization,
BSW matrix elements and the modelling of the absorptive contributions
in terms of penguin matrix elements. Factorization is receiving
increasing experimental support \cite{CLEO} and also most of the
BSW matrix elements have been tested experimentally in unsuppressed
decays \cite{Stone}. We are aware of the colour suppression problem
concerning the sign of $c_1+c_2/N$ \cite{CLEO,Stone}. But we prefer
to stay in line with the conventional QCD framework together with
factorization and the limit $N\to \infty$.

As in the case of $B\to VV$ decays, the inclusion of $1/N$ terms
would decrease the branching ratios for each of the channels involving
a $c\cb$-meson by about a factor of 25. Other channels which are drastically
effected by including $1/N$ terms are $B^- \to \pi^- \Phi$,
and $B^- \to K^- \omega$, where the branching ratios are decreased
by four orders of magnitude and a factor of about 20, respectively. In
all other channels, the effect of treating the $1/N$ terms amounts to
a change by less than a factor of two for the prediction of the branching
ratios or asymmetries (which generally change in the opposite direction).

A major simplification in our model is that we evaluated the matrix elements
at fixed $k^2=m_b^2/2$. This determines the imaginary parts of the
penguin diagrams. In a more elaborate treatment the effective $k^2$
would be determined by the dynamics. Here we relied on the work of
\cite{Deshpande} and \cite{SW} yielding a range of values given in
\eq(\ref{kk_range}) of which we have chosen the upper limit. To get
an idea how the asymmetries change with $k^2$ we
have calculated the asymmetries as a function of $k^2/m_b^2$. The
results are shown in Fig. 2, 3 and 4. Starting at zero for $k^2=0$
the asymmetries vary quite appreciably with $k^2$ (depending on the
thresholds appearing in the matrix elements), even when $k^2$
is restricted to the interval \eq(\ref{kk_range}). Only for the
transitions \xsixteen\ and \zfourteen\ the asymmetry is essentially
independent of $k^2$.
It is clear that the results presented in the tables in most cases are
sensitive to $k^2$. A more detailed investigation, channel by channel,
is needed to get a better estimate of the effective value of $k^2$, for
example, using the approach of Brodsky et al. \cite{BLS} which includes
extra gluon exchange or the model of Greub and Wyler \cite{Greub}
using wave functions.

In Tab.~1, 2, 3 and 4 we have given also results for some Cabibbo
allowed channels
$B^- \to K^- + \eta_c,\,D_s^- + D^0,\,D^- + D^0,\,K^- + J/\psi,\,
D^0 +D_s^{*-},\,K^{*-} + \eta_c,\,D^{*0} + D_s^-,\,D^0 + D^{*-}$ and
$D^{*0} + D^-$ for completeness. These results can be used for testing
some of our model assumptions. In these cases the asymmetries are not
significant except perhaps for $B^- \to D^0 + D^{*-}$ and $B^- \to D^{*0} +
D^-$. The channels with two charmed mesons in the final state could
also be calculated in the frame work of the heavy quark effective
theory which would be more rigorous than using BSW matrix elements.
However, it would give similar branching ratios \cite{KMP} and
asymmetries (assuming factorization and a fixed value of $k^2$).

We mention that some of the modes have been calculated by other authors
using a similar approach. First there is work on pure penguin modes
\cite{Flei,Fleis,Xing} and also some work on modes where penguins
interfere with tree diagrams \cite{Du}. Our results are in qualitative
agreement with these efforts. The work presented here is
more elaborate, because it includes NLO corrections in the $QCD$
coefficients for the tree diagrams and in the penguin contributions
which were not taken into account in \cite{Du}. Therefore it is not
too meaningful to make detailed comparisons with this work. The $k^2$
dependence of
asymmetries for some selected channels has been studied also
\cite{Du1,Du,Xing,Flei,Fleis} with similar results to ours.

The branching ratios we have presented in Tab.~1-4 are well within
experimental limits. However there are three channels where the most recent
limits \cite{Gronberg} come within factors of 2-3 of our predictions:
$K^0\pi^+,~K^{*0}\pi^+,~K^+\pi^0$. It is therefore clear that the next
generation of experiments can provide stringent tests of our model.

\subsection*{Acknowledgement}
W.\,F.\,P. thanks the Desy Theory Group for its kind hospitality and the North
Atlantic Treaty Organization for a Travel Grant. He is also grateful
for helpful conversations with K. Honscheid.

\section{Appendix}
\def\etau{\eta_u}
\def\etas{\eta_s}
The factorization coefficients $Z^{(q)}_{1,2}$ defined in
\eq(\ref{def_of_Z12}) are listed in Tab.~\ref{table_of_Z12} for
$B\to PP$ channels.
For the case of $B\to PV$ decays the factorization coefficients are
obtained by replacing one of the pseudoscalars by a vector meson with
the corresponding flavour content, and by setting $R[M,P] = 0$ if $M$
is a vector meson.

In Tab.~\ref{table_of_Z12} we refer to the unphysical states
\be
\etau \equiv \frac{1}{\sqrt{2}}(u\ub + d\db) \nonumber \ ,\quad \quad
\etas \equiv s\sb \ ,
\ee
which are related to the physical $\eta$-mesons by taking into account
corresponding mixing angles.
The table covers both cases, $q=u$ and $q=c$, of \eq(\ref{def_of_Z12});
by use of the Kronecker delta the contributions from the operators $O_{1,2}$
are to be dropped when tree contributions are absent for the given CKM
prefactor, see \eq(\ref{def_of_Hq}).
Colour suppressed terms may readily be included in the coefficients
\be
a_i \equiv c_i^\eff + \frac{1}{N} c_j^\eff
\label{def_of_ai} \ee
where $\{i,j\}$ is any of the pairs $\{1,2\}$, $\{3,4\}$, or $\{5,6\}$.
In Tab.~\ref{table_of_Z12} we have adopted the convention of including
factors of $\sqrt{2}$ associated with a neutral meson $P_2$. They arise
either from current matrix elements between $P_2$ and $B$ (left column),
or from the definition of the decay constants for $P_2$ (right column).
Care should be taken with the latter since these factors are sometimes
absorbed into the decay constants (e.g. as tabulated in \cite{BSW}).

\newpage
\section*{Figure captions}
\begin{description}

\item [Fig.\,1:] Two types of one-loop matrix elements: (a) Vertex
corrections, and (b) penguin diagrams. The square box denotes an
insertion of one of the four-quark operators $O_i$ of \eq(\ref{operators}).

\item [Fig.\,2:] Momentum dependence of the asymmetry parameter $a_{CP}$
for $B\to PP$ transitions and a $\rho$ {\it positive} CKM Matrix.

\item [Fig.\,3:] Momentum dependence of the asymmetry parameter $a_{CP}$
for $B\to PV$ ($b \to s$ transitions) and a $\rho$ {\it positive} CKM Matrix.

\item [Fig.\,4:] Momentum dependence of the asymmetry parameter $a_{CP}$
for $B\to PV$ ($b \to d$ transitions) and a $\rho$ {\it positive} CKM Matrix.
\end{description}

%\newpage
\section*{Table captions}
\begin{description}
\item [Tab.\,1:] Branching ratios and rate asymmetries for $B \rightarrow PP$
($b \rightarrow s$ transitions)
for a $\rho$ {\it positive} and a $\rho$ {\it negative} CKM Matrix, and
for $k^2=m_b^2/2$. Values in parentheses correspond to the case without
strong phases.

\item [Tab.\,2:] Branching ratios and rate asymmetries for $B \rightarrow PP$
($b \rightarrow d$ transitions)
for a $\rho$ {\it positive} and a $\rho$ {\it negative} CKM Matrix, and
for $k^2=m_b^2/2$. Values in parentheses correspond to the case without
strong phases.

\item [Tab.\,3:] Branching ratios and rate asymmetries for $B \rightarrow PV$
($b \rightarrow s$ transitions)
for a $\rho$ {\it positive} and a $\rho$ {\it negative} CKM Matrix, and
for $k^2=m_b^2/2$. Values in parentheses correspond to the case without
strong phases.

\item [Tab.\,4:] Branching ratios and rate asymmetries for $B \rightarrow PV$
($b \rightarrow d$ transitions)
for a $\rho$ {\it positive} and a $\rho$ {\it negative} CKM Matrix, and
for $k^2=m_b^2/2$. Values in parentheses correspond to the case without
strong phases.

\item [Tab.\,5:] Factorization coefficients $Z^{(q)}_{1,2}$ for various
$B\to P_1 P_2$ decays. The short distance coefficients $a_i$ are defined
in \eq(\ref{def_of_ai}) and the factor $R$ is given in \eq(\ref{def_of_R}).

\end{description}

\newpage

\def\tmtwo{$\times10^{-2}$}
\def\tmthree{$\times10^{-3}$}
\def\tmfour{$\times10^{-4}$}
\def\tmfive{$\times10^{-5}$}
\def\tmsix{$\times10^{-6}$}
\def\tmseven{$\times10^{-7}$}
\def\tmeight{$\times10^{-8}$}

% Spacings:
%\renewcommand{\arraystretch}{1.3}
\def\nc{\\}
\def\sc{\\[-1mm]}
\small

\begin{table}
\caption{~}
\begin{center}
\begin{tabular}{||l|c|c||c|c||}
\hline\hline
\multicolumn{5}{||c||}{$B \rightarrow PP$ ($b \to s$ transitions)} \\
\multicolumn{5}{||c||}{Matrix Elements {\it with (without)} Strong Phases
for $k^2=m_b^2/2$ }\\
\hline
CKM Matrix:&\multicolumn{2}{|c||}{$\rho=0.32, \eta=0.31$}&
                       \multicolumn{2}{|c||}{$\rho=-0.41, \eta=0.18$}\\
\hline
Channel   & BR  & $a_{CP} [\%]$& BR  & $a_{CP} [\%]$  \\
%\hline
%\multicolumn{5}{||c||}{$b \rightarrow s$ transitions:~~
%$\Delta c=0,~\Delta b =  \Delta s = 1 $}\\
\hline
\xone    & 1.4 \tmsix  & $-$3.2    &2.0 \tmseven& $-$12.  \\
\xtwo    & 1.5 \tmsix  &           &2.5 \tmseven&    \\
  &(1.4 \tmsix) &  &(2.2 \tmseven) & \\
\xthree  & 7.8 \tmsix   &  5.7     &1.0 \tmfive & 2.5   \\
\xfour   & 7.0 \tmsix   &          &9.4 \tmsix  &  \\
  &(6.6 \tmsix ) &  &(8.9 \tmsix ) & \\
\xfive   & 6.0 \tmsix  &  8.1      &1.1 \tmfive & 2.3   \\
\xsix     &5.1 \tmsix  &           &1.1 \tmfive &  \\
  &(5.3 \tmsix  ) &  &(1.1 \tmfive) & \\
\xseven       & 1.9 \tmthree  &  0.0  &1.9 \tmthree & 0.0  \\
  &(1.9 \tmthree) &  &(1.9 \tmthree) & \\
\xeight       & 7.1 \tmthree   &   $-$0.12 &7.1 \tmthree & $-$0.069 \\
\xeightb      & 7.1 \tmthree   &           &7.1 \tmthree &       \\
  &(7.1 \tmthree) &  &(7.1 \tmthree) & \\
\xseventeen   & 2.0  \tmfive   & 0.51    &1.9  \tmfive   & 0.31 \\
\xseventeenb  & 2.0  \tmfive   &         &1.9\tmfive     &\\
  &(1.9 \tmfive ) &  &(1.8 \tmfive )&  \\
\hline\hline
\end{tabular}
\end{center}
\end{table}

\eject

\begin{table}
\caption{~}
\begin{center}
\begin{tabular}{||l|c|c||c|c||}
\hline\hline
\multicolumn{5}{||c||}{$B \rightarrow PP$ ($b \to d$ transitions)} \\
\multicolumn{5}{||c||}{Matrix Elements {\it with (without)} Strong Phases
for $k^2=m_b^2/2$ }\\
\hline
CKM Matrix:&\multicolumn{2}{|c||}{$\rho=0.32, \eta=0.31$}&
                       \multicolumn{2}{|c||}{$\rho=-0.41, \eta=0.18$}\\
\hline
Channel   & BR  & $a_{CP} [\%]$& BR  & $a_{CP} [\%]$  \\
%\hline
%\multicolumn{5}{||c||}{$b \rightarrow d$ transitions:~~
%$\Delta s = \Delta c = 0,~\Delta b = 1 $}\nc
\hline
\xnine  & 6.6 \tmsix   & $-$13.   &2.1 \tmsix  &$-$21.\\
\xten   & 8.6 \tmsix   &          &3.2 \tmsix  & \\
  &(7.5 \tmsix) &  &(2.4 \tmsix)  & \\
\xeleven & 1.9 \tmfive  & $-$12.  &3.3 \tmfive  & $-$4.3\\
\xtwelve & 2.5 \tmfive  &         &3.6 \tmfive  & \\
  &(2.2 \tmfive) &  &(3.3 \tmfive) & \\
\xthirteen     & 5.6 \tmsix   &  0.0  &5.7 \tmsix  &  0.0 \\
%\xfourteen    & 5.6 \tmsix   &       &5.7 \tmsix  & \\
  &(5.6 \tmsix) &  &(5.7 \tmsix) & \\
\xfifteen  & 8.4 \tmfive &  0.0 &7.3 \tmsix  & 0.0 \\
  &(8.4 \tmfive) &  &(7.3 \tmsix) & \\
\xsixteen  & 4.0 \tmfour &  2.2    &3.4 \tmfour &  1.5  \\
\xsixteenb & 3.8 \tmfour &         &3.3 \tmfour &       \\
  & (3.9 \tmfour) &  &(3.4\tmfour ) & \\
\xeighteen   & 6.8  \tmseven  & $-$17.    &2.3  \tmsix    & $-$3.3 \\
\xeighteenb  & 9.4  \tmseven  &         &2.5\tmsix&     \\
  &(7.8 \tmseven )&  &(2.2 \tmsix  )&  \\
\hline\hline
\end{tabular}
\end{center}
\end{table}

\eject

\begin{table}
\caption{~}
\begin{center}
\begin{tabular}{||l|c|c||c|c||}
\hline\hline
\multicolumn{5}{||c||}{$B \rightarrow PV$ ($b \to s$ transitions) }\\
\multicolumn{5}{||c||}{Matrix Elements {\it with (without)} Strong Phases
for $k^2=m_b^2/2$}\\
\hline
CKM Matrix:&\multicolumn{2}{|c||}{$\rho=0.32, \eta=0.31$}&
                       \multicolumn{2}{|c||}{$\rho=-0.41, \eta=0.18$}\\
\hline
Channel   & BR  & $a_{CP}  [\%]$ & BR  & $a_{CP} [\%]$ \\
%\hline
%\multicolumn{5}{||c||}{$b \rightarrow s$ transitions:~~
%$\Delta c=0,~\Delta b =  \Delta s = 1 $}\\
\hline
\yone    & 7.4 \tmsix   &  1.0 &4.2 \tmsix    & 1.0  \\
\ytwo    & 7.3 \tmsix   &      &4.1 \tmsix    &   \\
  &(7.3 \tmsix) &  &(4.1 \tmsix  ) & \\
\ythree  & 2.7 \tmseven &  34.  &9.4 \tmseven & 4.4 \\
\yfour   & 1.4 \tmseven &       &8.6 \tmseven &  \\
  &(1.8 \tmseven) &  &(8.8 \tmseven)  & \\
\yfive    & 9.0 \tmfour   &  0.0  &9.1 \tmfour  &  0.0    \\
  &(9.0 \tmfour ) &  &(9.1 \tmfour ) & \\
\ysix     &8.8 \tmthree  & $-$0.047 &8.8 \tmthree  &$-$0.027  \\
\ysixb    &8.8 \tmthree  &          &8.8 \tmthree  &        \\
  &(8.8 \tmthree) &  &(8.8 \tmthree) & \\
\yseven       & 3.4 \tmseven &  45.  &9.9 \tmseven  & 6.7 \\
\yeight       & 1.3 \tmseven &       &8.7 \tmseven & \\
  &(2.2 \tmseven) &  &(9.2 \tmseven) & \\
\ynine  & 1.5 \tmsix   &  13.  &2.5 \tmseven & 64.  \\
\yten   & 1.2 \tmsix   &       &5.7 \tmeight & \\
        &(1.3 \tmsix  ) &  &(9.7 \tmeight) & \\
\yeleven   & 3.9 \tmsix  &  19.  &1.1 \tmfive  & 3.4 \\
\ytwelve & 2.6 \tmsix  &         &1.0 \tmfive  & \\
  &(3.1 \tmsix) &  &(1.1 \tmfive & \\
\ythirteen     & 5.8 \tmfour    &  0.0 &5.8 \tmfour & 0.0 \\
  &(5.8 \tmfour ) &  &(5.8 \tmfour ) & \\
\yfourteen    & 4.0 \tmthree   &$-$0.013 &4.0 \tmthree  &$-$0.007\\
\yfourteenb   & 4.0 \tmthree   &         &4.0 \tmthree  &       \\
  &(4.0 \tmthree) &  &(4.0 \tmthree) & \\
\yfifteen     & 1.5 \tmfive    &  0.54  &1.4 \tmfive   &  0.33 \\
\yfifteenb    & 1.5 \tmfive    &        &1.4 \tmfive   &       \\
  &(1.4 \tmfive ) &  &(1.3 \tmfive ) & \\
\ysixteen     & 5.8 \tmsix     &  1.2   &5.5 \tmsix    &  0.75 \\
\ysixteenb    & 5.7 \tmsix     &        & 5.4 \tmsix    &     \\
  &(4.5 \tmsix  ) &  &(4.3 \tmsix  ) & \\
\yseventeen     & 1.0 \tmsix     &  0.66  &9.8 \tmseven  &  0.41 \\
\yseventeenb    & 1.0 \tmsix     &        &9.7 \tmseven  &       \\
  &(9.4 \tmseven) &  &(8.9 \tmseven) & \\
\hline\hline
\end{tabular}
\end{center}
\end{table}

\eject

\begin{table}
\caption{~}
\begin{center}
\begin{tabular}{||l|c|c||c|c||}
\hline\hline
\multicolumn{5}{||c||}{$B \rightarrow PV$ ($b \to d$ transitions) }\\
\multicolumn{5}{||c||}{Matrix Elements {\it with (without)} Strong Phases
for $k^2=m_b^2/2$}\\
\hline
CKM Matrix:&\multicolumn{2}{|c||}{$\rho=0.32, \eta=0.31$}&
                       \multicolumn{2}{|c||}{$\rho=-0.41, \eta=0.18$}\\
\hline
Channel   & BR  & $a_{CP} [\%]$& BR  & $a_{CP} [\%]$  \\
%\hline
%\multicolumn{5}{||c||}{$b \rightarrow d$ transitions:~~
%$\Delta s =\Delta c=0,~\Delta b =   1 $}\\
\hline
\zone   & 7.8 \tmsix   & 6.5 & 1.8 \tmfive & 1.6 \\
\ztwo   & 6.8 \tmsix   &     & 1.7 \tmfive &  \\
 &(7.3 \tmsix ) &  &(1.8 \tmfive ) & \\
\zthree & 9.6 \tmsix   & 14.  &5.7 \tmfive  & 1.2 \\
\zfour  & 7.2 \tmsix   &      &5.6 \tmfive  & \\
 &(8.2 \tmsix ) &  &5.6 \tmfive) & \\
\zfive   & 2.3 \tmfive  & $-$2.0  &1.7 \tmfive &$-$1.5 \\
\zsix    & 2.4 \tmfive  &         &1.8 \tmfive &  \\
 &(2.3 \tmfive) &  &(1.8 \tmfive) & \\
\zseven  & 4.3 \tmfive   & 0.0 & 3.7 \tmfive & 0.0 \\
         & (4.3 \tmfive) &        &(3.7 \tmfive)&      \\
\zeight  & 5.0 \tmfour   & 0.89  &4.7 \tmfour &0.55 \\
\zeightb & 4.9 \tmfour   &       &4.7 \tmfour &\\
 &(5.0 \tmfour) &  &(4.7 \tmfour) & \\
\znine   & 1.4 \tmsix  & $-$15. & 1.3 \tmsix   & $-$9.7\\
\zten    & 1.8 \tmsix  &        & 1.6 \tmsix   & \\
 &(1.6 \tmsix) &  &(1.4 \tmsix  ) & \\
\zeleven  & 1.3 \tmsix  &  11.  &2.9 \tmsix & 2.5 \\
\ztwelve& 1.0 \tmsix    &       &2.8 \tmsix & \\
 &(1.2 \tmsix) &  &(2.9 \tmsix) & \\
\zthirteen & 2.7 \tmfive  &  0.0 &2.4 \tmfive   & 0.0 \\
           &(2.7 \tmfive) &      &(2.4 \tmfive) & \\
\zfourteen & 2.2 \tmfour&   0.23  &2.2 \tmfour& 0.14  \\
\zfourteenb & 2.2 \tmfour&         &2.2 \tmfour&       \\
 &(2.2 \tmfour &  &(2.2 \tmfour)  & \\
\zfifteen  & 1.1 \tmseven  & 0.0 &4.0 \tmseven   & 0.0 \\
           &(1.1 \tmseven) &     &(4.0 \tmseven) & \\
\zsixteen     & 3.5 \tmeight   & $-$21. &1.2 \tmseven  & $-$4.3 \\
\zsixteenb    & 5.4 \tmeight   &        &1.3 \tmseven  &        \\
  &(4.2 \tmeight) &  &(1.1 \tmseven) & \\
\zseventeen     & 4.7 \tmseven   & $-$18.   &1.6 \tmsix    & $-$3.5  \\
\zseventeenb    & 6.7 \tmseven   &        &1.7 \tmsix    &       \\
  &(5.5 \tmseven) &  &(1.5 \tmsix  ) & \\
\hline\hline
\end{tabular}
\end{center}
\end{table}

\begin{table}
\caption{~}
\begin{center}
\begin{tabular}{||c|c|c|c||}
\hline \hline
$P_1$ & $P_2$ & $Z^{(q)}_1$ & $Z^{(q)}_2$ \\ \hline
$K^-$ & $\pi^0$  & $(a_2 \d_{qu} + a_4 + a_6 \,R[K^-,\pi^0])/\sqrt{2}$
                 & $a_1 \d_{qu} /\sqrt{2}$ \\
$K^-$ & $\etau$ & $(a_2 \d_{qu} + a_4 + a_6 \,R[K^-,\eta_u])/\sqrt{2}$
                 & $(a_1 \d_{qu} + 2a_3 + 2a_5)/\sqrt{2}$ \\
$K^0$ & $\pi^-$  & $a_4 + a_6 \,R[K^0,\pi^-]$
                 & $0$ \\
$K^-$ & $\etas$ & 0
                 & $a_3 + a_4 + a_5 + a_6 \,R[\eta_s,K^-]$ \\
$D_s^-$ & $D^0$  & 0
                 & $a_2 \d_{qc} + a_4 + a_6 \,R[D_s^-,D^0]$ \\
$K^-$ & $\eta_c$ & 0
                 & $a_1 \d_{qc} + a_3 + a_5$ \\ \hline
$\pi^-$ & $\pi^0$& $(a_2 \d_{qu} + a_4 + a_6 \,R[\pi^-,\pi^0])/\sqrt{2}$
                 & $(a_1 \d_{qu} - a_4 - a_6 \,R[\pi^0,\pi^-])/\sqrt{2}$  \\
$\pi^-$ & $\etau$& $(a_2 \d_{qu} + a_4 + a_6 \,R[\pi^-,\eta_u])/\sqrt{2}$
                 & $(a_1 \d_{qu} + 2a_3 + a_4 + 2a_5 + a_6 \,R[\eta_u,\pi^-])
                   /\sqrt{2}$\\
$K^0$ & $K^-$  & $a_4 + a_6 \,R[K^0,K^-]$
                 & $0$ \\
$\pi^-$ & $\etas$ & 0
                 & $a_3 + a_5$ \\
$D^-$ & $D^0$    & 0
                 & $a_2 \d_{qc} + a_4 + a_6 \,R[D^-,D^0]$ \\
$\pi^-$ & $\eta_c$ & 0
                 & $a_1 \d_{qc} + a_3 + a_5$ \\ \hline
\hline
\end{tabular}
\label{table_of_Z12}
\end{center}
\end{table}

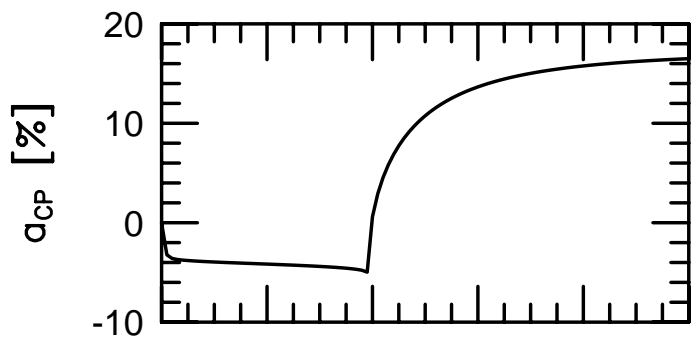
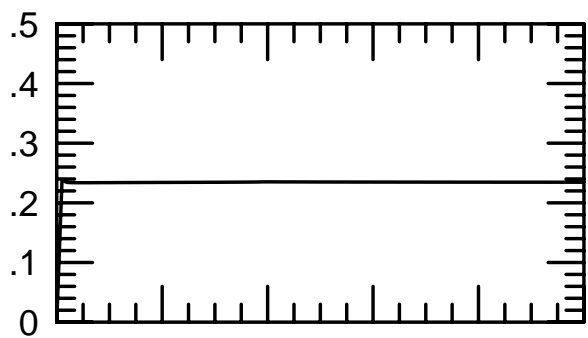
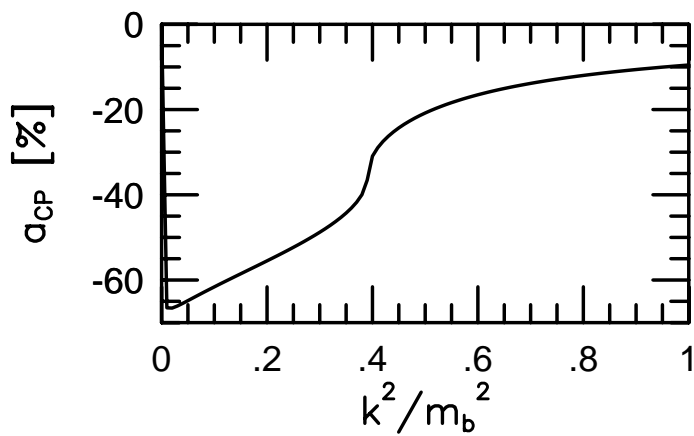
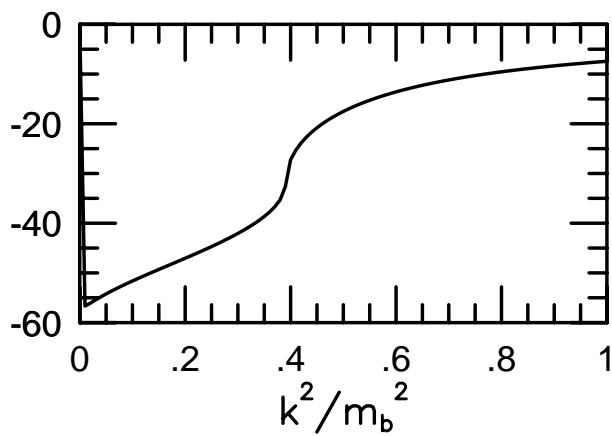
\end{document}